
\input harvmac.tex
\Title{\vbox{\baselineskip12pt\hbox{        }\hbox{}}}
{\vbox{{\centerline{Ferroelectric Phase Transitions}}
{\centerline{from First Principles}}}}
\bigskip
\centerline{K. M. Rabe and U. V. Waghmare}
\centerline{Department of Applied Physics}
\centerline{Yale University}
\centerline{New Haven, CT 06520}
\bigskip
\bigskip
\bigskip
\vskip 1in
\centerline{\bf Abstract}
\noindent
An effective Hamiltonian for the ferroelectric transition in $PbTiO_3$ is
constructed from first-principles density-functional-theory total-energy and
linear-response calculations through the use of a localized, symmetrized basis
set of ``lattice Wannier functions.''
Preliminary results of Monte Carlo simulations for this system show a
first-order cubic-tetragonal transition at 660 K.
The involvement of the Pb atom in the lattice instability and the coupling of
local distortions to strain are found to be particularly important in producing
the behavior characteristic of the $PbTiO_3$ transition.
A tentative explanation for the presence of
local distortions experimentally observed above $T_c$ is suggested.
Further applications of this method to a variety of systems and structures are
proposed for first-principles study of finite-temperature structural properties
in individual materials.
\vskip 0.2in
\noindent
{\it Presented at the Third Williamsburg Workshop on Fundamental Experiments in
Ferroelectrics, February 1995.}
\Date{}

\vfil\eject

\newsec{Introduction}
Structural phase transitions in perovskite structure compounds have long been
of experimental and theoretical interest
\ref\lg{M. E. Lines and A. M. Glass, {\it Principles and Applications of
Ferroelectrics and Related Materials} (Oxford, 1977), Chap. 8.}.
This class of materials exhibits a wide variety of transition behavior
and low temperature distorted structures, depending on the individual
compound.
{}From first-principles calculations, quantitative information on the
structural energetics can be obtained to understand these chemical trends.
Furthermore, this approach yields a microscopic description of the origin and
nature of lattice instabilities in these materials.

Success in applying the first-principles approach to perovskite oxides has been
achieved only rather recently, due to the presence of oxygen and transition
metal elements, the number of atoms in the unit cell, and the high accuracy
required for a meaningful computation of the delicate lattice instabilities.
The 1992 LAPW study
\ref\cohen{R. E. Cohen and H. Krakauer, Ferroelectrics {\bf 136}, 65 (1992); R.
E. Cohen, Nature {\bf 358}, 136 (1992).}
of $BaTiO_3$ and $PbTiO_3$ has been followed by investigations in a number of
systems using various implementations of density-functional-theory (DFT)
methods
\ref\kingsmith{R. D. King-Smith and D. Vanderbilt, Phys. Rev. {\bf B49}, 5828
(1994); W. Zhong, R. D. King-Smith and D. Vanderbilt, Phys. Rev. Lett. {\bf
72}, 3618 (1994).}
\ref\singh{D. J. Singh and L. L. Boyer, Ferroelectrics {\bf 136}, 95 (1992).}
\ref\rabe{K. M. Rabe and U. V. Waghmare, Ferroelectrics {\bf 151}, 59 (1994).}
\ref\mpost{M. Posternak, R. Resta and A. Baldereschi, Phys. Rev. {\bf B50},
8911 (1994).}
\ref\gonze{Ph. Ghosez, X. Gonze and J.-P. Michenaud, Ferroelectrics {\bf 153},
19 (1994).}
\ref\postni{A. V. Postnikov, T. Neumann and G. Borstel, Phys. Rev. {\bf B50},
758 (1994); A. V. Postnikov and G. Borstel, Phys. Rev. {\bf B50}, 16403
(1994).}
\ref\ky{H. Krakauer and R. Yu, to be published in {\it Phys. Rev. Lett.}}.
In this work, we combine total energy and electronic bandstructure calculations
with the DFT perturbation method
\ref\bgt{S. Baroni, P. Giannozzi and A. Testa, Phys. Rev. Lett. {\bf 58}, 1861
(1987); P. Giannozzi, S. de Gironcoli, P. Pavone and S. Baroni, Phys. Rev. {\bf
B43}, 7231 (1991).}
\ref\gat{X. Gonze, D. C. Allan and M. P. Teter, Phys. Rev. Lett. {\bf 68}, 3603
(1992).}
\ref\yale{U. V. Waghmare, V. Milman and K. M. Rabe, unpublished.}
\ref\krakauer{C. Z. Wang, R. Yu and H. Krakauer, Phys. Rev. Lett. {\bf 72}, 368
(1994); R. Yu and H. Krakauer, Phys. Rev. {\bf B49}, 4467 (1994).}.
The latter provides a means for computing force constant matrices at arbitrary
wavevectors $\vec q$ without the need for supercells,
thus facilitating the exploration of lattice instabilities throughout the first
Brillouin zone.
Furthermore, valuable information about polarization density in these systems
can be obtained from direct computation of the Born effective charge tensors
and electronic dielectric constant $\epsilon_\infty$.

The highly demanding nature of these computations has the consequence that
direct molecular dynamics or Monte Carlo simulations of temperature-driven
structural transitions in perovskite oxides using DFT forces and total energies
is completely impractical.
Indeed, the investigations mentioned above do not attempt to map out
a full configurational energy surface, but rather focus on those aspects felt
to be most relevant to the description of the structural transitions:
(i) lattice instabilities at quadratic order;
(ii) anharmonicity at $\vec q = 0$;
 and (iii) coupling of unstable modes to strain.
We have developed a method based on the construction of a ``lattice Wannier
function'' basis
\ref\lwf{K. M. Rabe and U. V. Waghmare, mtrl-th 9411006.}
which uses this first-principles information for an individual material
efficiently to derive an effective Hamiltonian in the low-energy subspace of
the ionic displacement space relevant to the temperature range of the
transition.
This effective Hamiltonian can be regarded as a realistic generalization of the
$\Phi^4$ models that are widely used in studies of generic ferroelectric
transition behavior
\ref\fahy{S. Fahy and R. Merlin, Phys. Rev. Lett. {\bf 73}, 1122 (1994).}
\ref\heine{A. P. Giddy, M. T. Dove and V. Heine, J. Phys. Cond. Matt. {\bf 1},
8327 (1989); B. G. A. Normand, A. P. Giddy, M. T. Dove and V. Heine, J. Phys.
Cond. Matt. {\bf 2}, 3737 (1990)}
\ref\pad{S. Padlewski, A. K. Evans, C. Ayling and V. Heine, J. Phys. Cond.
Matt. {\bf 4}, 4895 (1992).}
A Monte Carlo simulation of this effective Hamiltonian, by construction, will
quantitatively reproduce the $T\neq 0$ behavior of the particular system under
consideration.

To illustrate this approach, we present preliminary results of its application
to the ferroelectric transition in $PbTiO_3$.
This compound has the cubic perovskite structure at high temperatures,
undergoing a first-order transition at 763 K to a tetragonal phase produced by
relative displacement of the cationic and oxygen sublattices resulting in a
nonzero uniform polarization density.
In the first-principles study of this transition, we obtain results for the
transition temperature, the order of the transition and the nature of the
paraelectric phase.
The role of different contributions to the effective Hamiltonian, such as
strain coupling and the $\vec q$ dependence of the lattice instability, can be
examined.
The significant differences from $BaTiO_3$
\ref\zvr{W. Zhong, D. Vanderbilt and K. M. Rabe, Phys. Rev. Lett. {\bf 73},
1861 (1994); {\it ibid.}, unpublished.} and $SrTiO_3$
\ref\zvs{W. Zhong and D. Vanderbilt, Phys. Rev. Lett. {\bf 74}, 2587 (1995).}
become evident.
Although $PbTiO_3$ has been described as a ``textbook example of
a displacive ferroelectric transition,'' \lg~ recent experimental evidence and
our first-principles results suggest that the apparent simplicity of this
behavior is quite nontrivial.

\newsec{Derivation of the Effective Hamiltonian}
The derivation of the effective Hamiltonian \lwf~is here summarized briefly.
The starting point of the analysis is the classical partition function
\eqn\part{Z \propto \int\{d \vec u_j\} exp(-\beta \CH_{lat}(\{\vec u_j\}))}
where, as a result of the Born-Oppenheimer approximation, only the ionic
degrees of freedom $\{\vec u_{j}\}$ appear explicitly, the index $j$ running
over all ions in the system.
The Taylor expansion of the full lattice Hamiltonian around a high-symmetry
reference structure, such as the cubic perovskite structure, can be grouped
into the harmonic terms $\CH_h$ and higher-order terms in the displacements
$\CH_{anh}$.
Diagonalization of the harmonic part $\CH_h$ is straightforward, and the
results can be depicted using the dispersion relation of the type shown in
\fig\disp{A sketch showing the features of the dispersion relation
$\omega^2_{\lambda}(\vec q)$ characteristic of systems undergoing structural
transitions, and a possible decomposition into the two subspaces $\Lambda_s$
and $\Lambda_0$ described in the text.}.
This suggests a natural decomposition of the full ionic displacement space into
two subspaces:
(i) $\Lambda_0$, composed of the branches of the dispersion relation that
contain unstable modes characteristic of the structural transition,
and
(ii) $\Lambda_s$, composed of the remaining branches, for which typically all
modes have $\omega^2(\vec q)\ge 0$.
This decomposition permits the following approximation to the lattice
Hamiltonian:
\eqn\decoup{\CH_{lat}\approx
\CH_{h}(\Lambda_0)+\CH_{anh}(\Lambda_0)+\CH_{h}(\Lambda_s)}
where only the anharmonic terms in the unstable subspace $\Lambda_0$, crucial
for stabilizing the crystal, are included.
As a result, the $\Lambda_s$ integral in the partition function becomes
trivial, making no contribution to the temperature dependence of the average
structure, and the partition function reduces to
$Z\approx\int_{\Lambda_0}exp(-\beta\CH_{eff}(\Lambda_0))$
where
$\CH_{eff}(\Lambda_0)\equiv
\CH_{h}(\Lambda_0)+\CH_{anh}(\Lambda_0)$.

An explicit form for the effective Hamiltonian $\CH_{eff}(\Lambda_0)$ is
obtained by making a Taylor expansion in coordinates of the subspace
$\Lambda_0$.
The simplicity of this expression can be maximized by a proper choice of the
corresponding basis.
In analogy to the tight-binding representation of a subset of electronic
eigenstates in a crystal, we choose a highly symmetric localized basis for
$\Lambda_0$, where each basis vector $w_{i\lambda}$ is associated with a
Wyckoff position in the unit cell $\vec R_i$ and transforms according to an
irreducible representation of its site symmetry group, the Wyckoff positions
and irreps being determined by the space group irrep labels of $\Lambda_0$.
These localized basis functions can be related to the Bloch vectors $e_{\vec q
\lambda}$, which diagonalize $\CH_{h}(\Lambda_0)$, through an expression of the
form
$w_{i\lambda}=\sum_{\vec q}e^{i\vec q \cdot \vec R_i}e_{\vec q \lambda}$ and,
by analogy with the electronic case, are given the name ``lattice Wannier
functions.''
With this basis, an ionic configuration in the subspace $\Lambda_0$ can be
specified by the corresponding coordinates $\{\xi_{i\lambda}\}$.

\newsec{Effective Hamiltonian for $PbTiO_3$}

The first-principles method for the construction of the effective Hamiltonian
for an individual material proceeds by the following steps.
(1) The subspace $\Lambda_0$ is identified, a symmetry analysis performed, and
approximate Wannier basis vectors constructed by fitting to force-constant
matrix eigenvectors at high-symmetry $\vec q$-points.
(2) The form for the effective Hamiltonian is obtained by a Taylor expansion in
symmetry-invariant combinations of the Wannier coordinates
$\{\xi_{i\lambda}\}$. The expansion is truncated to include a relatively small
number of physically important terms.
(3) Finally, the set of parameters is  determined from first principles by
fitting $\CH_{eff}$ to the results of selected total energy and linear response
calculations, using the explicit correspondence, obtained from the construction
in step (1), between the Wannier coordinate description $\{\xi_{i\lambda}\}$
and the actual ionic displacements $\{\vec u_{j}\}$. For each material,
overdetermination of the parameters through additional independent
first-principles calculations is recommended to establish the validity of the
truncations made in step (2).

\subsec{Construction of Wannier basis}

The relevant mode for the ferroelectric transition in $PbTiO_3$, which freezes
in to produce the low-temperature tetragonal structure, is the three-fold
degenerate zone-center $\Gamma_{15}$ mode.
For a complete identification of the unstable subspace $\Lambda_0$, DFT linear
response calculations were carried out at $\Gamma$, $R$, $X$ and $M$ (in the
Brillouin zone of the simple cubic lattice)
\ref\pbtio{K. M. Rabe and U. V. Waghmare, unpublished.}.
The resulting low energy eigenmodes $\Gamma_{15}$,
$R_{15}$, $X_5'$, $M_2'$ and $M_5'$ require that the Wannier basis vectors
$w_{i\lambda},~\lambda=1,2,3$ are situated at the Wyckoff position occupied by
the Pb atom and transform according to the vector irrep of the site symmetry
group $O_h$. This results from the significant involvement of the Pb atom in
the lattice instability
\ref\wburgold{K. M. Rabe and U. V. Waghmare, Ferroelectrics {\bf 164}, 15
(1995).} and deviates from what would be expected if the transition could be
viewed as resulting from the ``rattling'' Ti ion picture developed by Slater
for $BaTiO_3$
\ref\rattle{J. C. Slater, Phys. Rev. {\bf 78}, 748 (1950).}, in which case the
Wannier basis vectors would be situated at the Ti atom positions.

Following the procedure described in detail in Ref. \lwf, the Wannier basis
vectors are parametrized by identifying the displacement patterns centered at
the Pb atoms which transform according to the vector irrep of $O_h$, starting
from the innermost coordination shells.
In the present construction, we include displacements of the central, first and
second neighbor Pb ions and the nearest Ti and O shells, for a total of nine
parameters
(\fig\pattern{Independent displacement patterns for the innermost coordination
shells
of the lattice Wannier functions transforming like the $\hat z$ vector
component at the Pb atom positions. The open squares represent the Pb ions, the
shaded squares the Ti ions, and the open and shaded circles represent
inequivalent oxygen ions.
The three additional displacement patterns for first and second neighbor Pb
shells, included in the analysis, are not shown. The $\hat x$- and $\hat
y$-like components are obtained by rotating the patterns shown by ${\pi \over
2}$ around an appropriate cartesian axis.}).
The values obtained by fitting to the normalized eigenvectors calculated at
$\Gamma_{15}$, $R_{15}$, $M_2'$ and $M_5'$ show a rapid decay with increasing
distance from the central atom, as expected.
Futhermore, the $X_5'$ eigenvector predicted from the parameters obtained from
the fit at the other high-symmetry points is in excellent agreement with the
normalized $X_5'$ eigenvector calculated from first principles.
The results are given in detail in Ref. \pbtio.

\subsec{Form of the effective Hamiltonian}

With this choice of basis, the system is described as a set of
three-dimensional vectors $\{\vec \xi_i\}$ at the sites of a simple cubic
lattice, and
$\CH_{eff}$ is expanded in symmetry-invariant combinations with respect to
$O_h^1$, the space group of the cubic perovskite structure.
In the present work, intercell interactions are included up to quadratic order
only.  For first, second and third neighbors, the most general quadratic
interactions allowed by symmetry are included:
\eqn\first{\sum_i\sum_{\hat d=nn1}[
a_L(\vec\xi_i\cdot \hat d)(\vec\xi_i(\hat d)\cdot \hat d)
+a_T(\vec\xi_i\cdot\vec\xi_i(\hat d)
-(\vec\xi_i\cdot \hat d)(\vec\xi_i(\hat d)\cdot \hat d))]}
\eqn\second{+\sum_i\sum_{\hat d=nn2}[
b_L(\vec\xi_i\cdot \hat d)(\vec\xi_i(\hat d)\cdot \hat d)
+b_{T1}(\vec\xi_i\cdot \hat d_1)(\vec\xi_i(\hat d)\cdot \hat d_1)
+b_{T2}(\vec\xi_i\cdot \hat d_2)(\vec\xi_i(\hat d)\cdot \hat d_2)]}
\eqn\third{+\sum_i\sum_{\hat d=nn3}[
c_L(\vec\xi_i\cdot \hat d)(\vec\xi_i(\hat d)\cdot \hat d)
+c_T(\vec\xi_i\cdot\vec\xi_i(\hat d)
-(\vec\xi_i\cdot \hat d)(\vec\xi_i(\hat d)\cdot \hat d))],}
while beyond third neighbor we use a dipole-dipole form parametrized by the
mode effective charge $\overline Z^*$ and the electronic dielectric constant
$\epsilon_\infty$:
\eqn\dipole{\sum_i\sum_{\vec d}{(\overline Z^*)^2 \over \epsilon_\infty}
{(\vec\xi_i\cdot\vec\xi_i(\hat d)-
3(\vec\xi_i \cdot \hat d)(\vec\xi_i(\hat d)\cdot \hat d))\over |\vec d|^{3}}.}
Terms in the onsite potential, depending only on values of $\vec \xi_i$ at a
single $i$, include isotropic terms up to eighth order in $\vert \vec
\xi_i\vert$
and full cubic anisotropy at fourth order:
\eqn\onsite{\sum_i(A|\vec \xi_i |^2+B|\vec \xi_i |^4+
C(\xi_{ix}^4+\xi_{iy}^4+\xi_{iz}^4)
+D|\vec \xi_i |^6+E|\vec \xi_i |^8).}
The energy associated with homogeneous strain, specified by the tensor
$e_{\alpha\beta}$, $(\alpha,~\beta=x,y,z)$, and its coupling to the local
distortion $\vec \xi_i$, is included to lowest nontrivial order:
$$+{N\over 2}C_{11}\sum_\alpha e_{\alpha \alpha}^2
+{N\over 2}C_{12}\sum_{\alpha \neq \beta}e_{\alpha \alpha}e_{\beta \beta}
+{N\over 4}C_{44}\sum_{\alpha \neq \beta}e_{\alpha \beta}^2
+Nf\sum_\alpha e_{\alpha \alpha}$$
$$+g_0(\sum_\alpha e_{\alpha \alpha})\sum_i|\vec \xi_i |^2
+g_1\sum_\alpha(e_{\alpha \alpha}\sum_i\xi_{i \alpha}^2)
+g_2\sum_{\alpha < \beta}e_{\alpha \beta}\sum_i\xi_{i \alpha}\xi_{i \beta}$$
For our calculations in $PbTiO_3$, we take $e_{\alpha\beta}=0$ at $a_0$=3.96883
\AA, the lattice constant  measured for the cubic phase just above $T_c$.
We have further generalized this expression to include inhomogeneous strain by
expanding the effective Hamiltonian subspace according to the procedure
described in \lwf. The explicit expression is given in \pbtio.

\subsec{Determination of parameters from first principles}

With the truncations described in the previous section, the effective
Hamiltonian for $PbTiO_3$ contains a total of 21 parameters to be determined
from first-principles calculations.
These are performed with the preconditioned conjugate-gradients
plane-wave pseudopotential method, using the program CASTEP 2.1
\ref\payne{M. C. Payne, D. C. Allan, T. A. Arias, M. P. Teter and J. D.
Joannopoulos, Rev. Mod. Phys. {\bf 64}, 1045 (1992).},
and with the variational DFT perturbation method, using a program based on the
formalism in \gat.
Full details of the calculations are provided elsewhere
\pbtio.

The determination of the quadratic coupling parameters is greatly facilitated
by the use of the DFT perturbation method for the direct calculation of force
constant matrices at arbitrary $\vec q$, circumventing the need for
computationally intensive supercell calculations. In addition, this method can
be used for the direct calculation of Born effective charges $Z^*$ and the
dielectric constant $\epsilon_\infty$, from which the dipole-dipole term in
$\CH_{eff}$ is immediately obtained.
At a given $\vec q$, the energies of the three eigenvectors of the quadratic
part of the effective Hamiltonian are obtained by computing the quadratic part
of the first-principles total energy of the ionic displacement patterns in the
$\Lambda_0$ subspace from the DFT force constants at that $\vec q$.
The eight short-range intercell interactions $a_L$, $a_T$, $b_L$, $b_{T1}$,
$b_{T2}$, $c_L$ and $c_T$ are then obtained by requiring that
the expressions for the effective Hamiltonian eigenenergies in terms of $Z^*$,
$\epsilon_\infty$ and the short range intercell interactions  reproduce these
first-principles values for selected modes at a set of $\vec q$.
This process is illustrated in
\fig\dispersion{Determination of quadratic short-range intercell interaction
parameters for $PbTiO_3$. The solid and open circles indicate the
first-principles values for the energies of effective Hamiltonian eigenvectors.
The solid circles are used to determine values for the short-range interaction
parameters, which are used to generate the effective Hamiltonian dispersion
curves, shown as solid lines.
The open circles are first-principles results not included in the parameter
determination, for comparison.}.
The validity of the dipole-dipole form for intercell interactions beyond third
neighbor can examined by comparing the values predicted by $\CH_{eff}$ for
calculated energies not included in the determination of the parameters. In
\dispersion, these comparisons for $\vec q$ along (111) are shown to be very
good.

Given the quadratic coupling parameters, the higher order terms in the onsite
potential ($B$, $C$, $D$ and $E$), the elastic constants ($C_{11}$, $C_{12}$
and $C_{44}$) and the couplings between strain and local distortions ($g_0$,
$g_1$ and $g_2$) can be readily determined from the energies of configurations
in which $\vec \xi_i$ is independent of $i$, requiring only total energy
calculations for primitive unit cells containing five atoms. Further details of
this procedure and the complete set of numerical parameter values for the
$PbTiO_3$ effective Hamiltonian are given in \pbtio.

\newsec{Monte Carlo Study of the $PbTiO_3$ transition}

The form of the effective Hamiltonian, while greatly simplified relative to the
original lattice Hamiltonian, is still sufficiently complicated to discourage
the application of analytical statistical mechanics methods such as
renormalization group analysis or high-temperature expansions. However, it is
quite suitable for Monte Carlo simulation, since changes in energy with
configuration are readily calculated numerically
\ref\allen{M. P. Allen and D. J. Tildesley, {\it Computer Simulation of
Liquids}, (Oxford, 1987), Chap. 4.}.
We use a single-flip Metropolis approach, with runs ranging between 25,000 and
150,000 sweeps through the lattice.
Periodic boundary conditions are imposed on an $L\times L\times L$ simulation
cell, with finite-size scaling applied for $L$ ranging from 5 to 11.

In a finite-size simulation, a first-order transition such as the ferroelectric
transition in $PbTiO_3$ leads to the coexistence of two distinct phases,
separated by an energy barrier, in a temperature range near $T_c$.
Recently developed Monte Carlo methods for first-order transitions
\ref\firstord{B. A. Berg and T. Neuhaus, Phys. Rev. Lett. {\bf 68}, 9 (1992);
C. Borgs and W. Janke, Phys. Rev. Lett. {\bf 68}, 1738 (1992); W. Janke, Phys.
Rev. {\bf B47}, 14757 (1992).} determine the transition temperature from the
condition that the difference in the free energies of the two phases be zero.
If the two phases are sampled ergodically in the simulation, this difference is
quite easy to compute.
However, in cases of strongly first-order transitions, unbiased sampling of the
two phases can be extremely difficult to achieve, especially as the
simulation-cell size increases.
While we are presently considering various algorithms suitable for this
situation,
for now we put bounds on $T_c$ by monitoring the sensitivity of the average
structural parameters to the choice of initial state: $T_>$ is the lowest
temperature at which the
system averages are characteristic of the cubic state, starting with an inital
ground state tetragonal configuration, while $T_<$ is the highest temperature
at which a starting cubic configuration results in a tetragonal state. These
bounds are plotted in
\fig\tc{Monte Carlo estimate of $T_c$ as a function of increasing
simulation-cell size $L$. At each $L$, the vertical line extends from the lower
bound $T_<$ to the upper bound $T_<$, calculated as described in the text.}.
A value of $T_c=660 K$, obtained from averaging the bounds at the largest
system size, is in very good agreement with the experimental transition
temperature 763K. An estimate of a latent heat of 3400 J/mol, extracted from
separate runs for the two phases, obtained by using different inital states at
$T_c$, is in rough agreement with the measured value of 4800 J/mol
\ref\latentpb{G. Shirane and E. Sawaguchi, Phys. Rev. {\bf 81}, 458 (1951).},
and very much larger than the 209 J/mol latent heat of the cubic-tetragonal
transition in $BaTiO_3$ \ref\latentba{G. Shirane and A. Takeda, J. Phys. Soc.
Jpn. {\bf 7}, 1 (1952).}.

One of the opportunities offered by this first-principles analysis is to
investigate the role of different contributions in determining the behavior at
the transition.
One particularly striking result of this procedure is that if the strain
coupling parameters $g_0$, $g_1$ and $g_2$ are set to zero, the character of
the transition changes significantly.
Specifically, the system is then found to undergo a {\it second-order}
transition from the high-temperature cubic phase to a low-temperature {\it
rhombohedral} phase with a lower transition temperature of 400K.
Thus, not only is the strain coupling responsible for stabilizing the
ground-state tetragonal structure, as has been previously noted \cohen, but it
is also the key factor in producing the strong first-order character of the
transition.

Another advantage of computer simulation of a realistic first-principles model
is access to microscopic information about short-range order.
This is particularly of interest for the high-temperature paraelectric phase. A
microscopically nonpolar character for this phase, where the probability
distribution of local distortions has a single peak centered at zero, is
generally associated with a displacive nature for the transition.
In general, a single-well potential will result in a single-peak distribution
if intercell correlations are unimportant, as in the limit of high
temperatures. A connection between a single-well onsite potential and
displacive behavior has been further supported by a dynamical study of a
$\Phi^4$ model \pad.

At first glance, the situation in $PbTiO_3$ appears to be straightforward. The
onsite potential has a single-well character (positive quadratic coefficient in
\onsite), which by the above arguments seems to explain the reported displacive
character of the $PbTiO_3$ transition \lg.
However, there is recent evidence that this transition cannot be simply
described as displacive.
Recent EXAFS measurements
\ref\sicron{N. Sicron, B. Ravel, Y. Yacoby, E. A. Stern, F. Dogan and J. J.
Rehr, Phys. Rev. {\bf B50}, 13168 (1994).} show that local distortions,
characteristic of the low-temperature structure, persist at least as high as
85K above $T_c$.
In our theoretical approach, we can suggest a possible explanation. In  the
realistic $H_{eff}$, unlike the $\Phi^4$ model, we find that unstable modes of
the perovskite structure extend all the way to the zone boundary (\dispersion).
The associated lowering of the energy for local distortions, even in the
absence of long range order, could significantly affect the short-range order
above $T_c$. This feature, rather than the sign of the quadratic coefficient in
the onsite potential, may be the signature of a microscopically polar
paraelectric phase.
More detailed Monte Carlo simulations, including the effective Hamiltonian
generalized to include inhomogeneous strain, are in progress. In addition,
effective-Hamiltonian molecular dynamics studies are planned which could
reconcile large local distortions with the observed ``displacive'' behavior of
the soft mode above $T_c$.

\newsec{Further applications}

The lattice Wannier function method for the construction of effective
Hamiltonians is applicable to finite-temperature structural properties in a
variety of systems.
Effective Hamiltonian studies of structural phase transitions currently in
progress include the antiferroelectric transition in $PbZrO_3$
\ref\pbzro{U. V. Waghmare and K. M. Rabe, in preparation.},  the ferroelectric
transition in $KNbO_3$
\ref\knbo{R. Yu, U. V. Waghmare, H. Krakauer and K. M. Rabe, in preparation.},
and the cubic-tetragonal transition in $ZrO_2$ \ref\zro{R. B. Phillips, U. V.
Waghmare, and K. M. Rabe, in preparation.}.
Effective Hamiltonians can also be constructed to model the temperature
dependence of structural parameters in systems like pyroelectric $BeO$
\ref\alex{B. A. Elliott, U. V. Waghmare and K. M. Rabe, in preparation.}.
Because of the simplicity of the kinetic energy in a Wannier basis constructed
from the dynamical matrix eigenvectors, this approach is readily extended
beyond the equilibrium classical statistical mechanics analysis we have
presented here.
For example, it is straightforward to construct a quantum mechanical model, for
example, for a quantum paraelectric such as $SrTiO_3$
\ref\martonak{R. Martonak and E. Tosatti, Phys. Rev. {\bf B49}, 12596 (1994).}.
In addition, important issues in the dynamical properties of ferroelectrics,
such as soft-mode behavior and switching, can be studied in molecular dynamics
simulations for individual materials.

\newsec{Summary and Conclusions}

In conclusion, we have described a systematic method for the construction of
effective lattice Hamiltonians from first-principles calculations through the
use of a localized, symmetrized basis set of ``lattice Wannier functions.''
Applying this method to the ferroelectric transition in $PbTiO_3$, we find a
first order cubic-tetragonal transition at 660 K.
The involvement of the Pb atom in the lattice instability and the coupling of
local distortions to strain are found to be particularly important in producing
the behavior characteristic of the $PbTiO_3$ transition. A tentative
explanation for the presence of
local distortions experimentally observed above $T_c$ is suggested.
Further applications of this method to a variety of systems and structures are
proposed for first-principles study of finite-temperature structural properties
in individual materials.

\vskip 0.2in
\centerline{\bf Acknowledgments}
We are grateful for useful discussions with Volker Heine, Ronald Cohen, Henry
Krakauer, Rici Yu, Cheng-Zhang Wang, Ekhard Salje,  Clive Randall, Weiqing
Zhong and David Vanderbilt. We thank M. C. Payne and V. Milman for the use of
and valuable assistance with CASTEP 2.1. This work was supported by ONR Grant
N00014-91-J-1247. In addition, K. M. R.  acknowledges the support of the Clare
Boothe Luce Fund and the Alfred P. Sloan Foundation.

\listrefs

\listfigs

\bye